\documentclass[sigconf]{acmart}

\AtBeginDocument{%
}

\setcopyright{none}
\renewcommand\footnotetextcopyrightpermission[1]{}
\copyrightyear{2026}
\acmYear{2026}
\acmDOI{}
\acmConference[KDD EvalTrust Workshop 2026]{KDD Workshop on Evaluation and Trustworthiness of Agentic AI}{August 2026}{Jeju, Republic of Korea}
\acmISBN{}

\title[Evaluating Enterprise Analytics Agents]{Evaluating Enterprise Analytics Agents: An End-to-End, Trace-Backed Methodology}

\definecolor{addblue}{RGB}{25,60,130}

\author{Teja Venkat Kolli}
\email{tkolli@thumbtack.com}
\affiliation{%
  \institution{Thumbtack}
  \city{San Francisco}
  \state{CA}
  \country{USA}
}

\author{Sang Su Lee}
\email{psulee@thumbtack.com}
\affiliation{%
  \institution{Thumbtack}
  \city{San Francisco}
  \state{CA}
  \country{USA}
}

\author{Xueying Yan}
\email{xyan@thumbtack.com}
\affiliation{%
  \institution{Thumbtack}
  \city{San Francisco}
  \state{CA}
  \country{USA}
}

\author{Jessie Chen}
\email{jechen@thumbtack.com}
\affiliation{%
  \institution{Thumbtack}
  \city{San Francisco}
  \state{CA}
  \country{USA}
}

\author{Chi Cheng}
\email{ccheng@thumbtack.com}
\affiliation{%
  \institution{Thumbtack}
  \city{San Francisco}
  \state{CA}
  \country{USA}
}

\author{Kartik Ravisankar}
\email{kravisankar@thumbtack.com}
\affiliation{%
  \institution{Thumbtack}
  \city{San Francisco}
  \state{CA}
  \country{USA}
}

\author{Shishir Dash}
\email{shishirdash@thumbtack.com}
\affiliation{%
  \institution{Thumbtack}
  \city{San Francisco}
  \state{CA}
  \country{USA}
}

\author{Vijay Anand Raghavan}
\email{vraghavan@thumbtack.com}
\affiliation{%
  \institution{Thumbtack}
  \city{San Francisco}
  \state{CA}
  \country{USA}
}

\begin{document}

\begin{abstract}
Enterprise analytics agents are not only text-to-SQL systems. They interpret
business intent and choose metric definitions. They select data sources, execute
tools, inspect results, and produce natural-language answers. Those answers may
influence operational, financial, or executive decisions. Grading final answers
hides where these agents fail. A plausible answer can use the wrong source of
truth. It can skip a required decomposition, claim causality without support, or
change its table interpretation across repeated runs.

We present an end-to-end evaluation methodology for analytics agents. The
methodology grades agent behavior across three families: semantic understanding,
execution quality, and reliability. Grading uses question banks with
human-written golden answers, repeated runs, and runtime traces. Each run first
passes a run-validity check, then receives tiered, abstention-aware scores that
feed a decision framework rather than a release gate.

We instantiate the methodology on a controlled internal analytics agent at a
large online marketplace. The case study uses 50 analytics questions, two
anonymous model configurations, and three randomized repetitions per
configuration, yielding 300 traces. The higher-capability configuration reduced
early refusal from 73\% to 0\% and increased real-data answers from 21\% to
73\%. However, it also exhausted the tool-round budget on 16\% of runs. It
overran the schema-exploration budget on 77\% of traces. It changed its table
interpretation on 41 of 50 questions. On finance questions with structured
golden answers, source table use and escalation improved, but canonical
decomposition remained weak in both configurations. These results show why
trust in analytics agents requires end-to-end, trace-backed evaluation rather
than SQL correctness or final answer quality alone.
\end{abstract}




\maketitle

\section{Introduction}

Analytics agents are becoming natural-language interfaces to enterprise data
warehouses. A user may ask why revenue changed. Another user may ask whether a
daily metric is unusual or what is driving a forecast gap. To answer, an agent
must interpret the business question. It must map that question to the correct
metric definition, source of truth, query, and final response.

This is an analytical workflow, not a single generation task. A final answer
can look reasonable while being wrong for many reasons. The agent may choose an
off-source table or use the wrong date window. It may miss a business-rule
filter, skip a required decomposition, or narrate unsupported causal claims.
Conversely, an answer may be numerically close while failing the analytical
contract that a domain expert would require.

The evaluation problem is therefore broader than SQL accuracy. SQL correctness
is necessary, but it is not sufficient. Trustworthy analytics agent evaluation
must inspect how the answer was produced. It must check whether the agent
followed the expected method. It must also test whether repeated runs produce
a stable interpretation.

We propose an end-to-end evaluation methodology for decision-grade analytics
agents. We use ``decision-grade'' narrowly: evaluation should show whether an
agent answer is reliable enough to inform a business decision. It should also
show whether the agent should abstain, escalate, or route the answer to human
review. We position the methodology as a measurement foundation for agent
governance, not as a substitute for it.

We make three contributions:

\begin{enumerate}
  \item We define a workflow-level methodology for analytics-agent evaluation.
  It covers interpretation, source selection, SQL and tool execution,
  analytical method, grounding, confidence, and repeatability.
  \item We provide an analytics-specific failure taxonomy. It connects generic
  trajectory failures to concrete data and decision risks.
  \item We provide an empirical demonstration on a 300-trace case study
  (Section~\ref{sec:case}). It
  shows that a higher-capability configuration can answer more questions
  while still failing source selection, decomposition, grounding, efficiency,
  and cross-run consistency.
\end{enumerate}

\section{Why Analytics-Agent Evaluation Is Different}
\label{sec:why}

Standard text-to-SQL evaluation asks whether a query returns the expected
answer for a natural-language question. Enterprise analytics agents face a
larger task. They must reason about ambiguous business language, local metric
definitions, source of truth hierarchy, table grain, and time windows. They
must also reason about the decision the user is trying to make.

Our methodology starts from four human-review dimensions:

\begin{itemize}
  \item \textbf{Question interpretation.} Did the agent understand the metric,
  scope, time range, segment, and decision context?
  \item \textbf{SQL and data sourcing.} Did the agent use the right tables,
  joins, filters, grains, and calculations?
  \item \textbf{Insight reasoning.} Did the agent interpret the results,
  decompose drivers, and avoid unsupported causal claims?
  \item \textbf{Output quality.} Was the answer complete, readable, qualified,
  and useful to the user?
\end{itemize}

These four dimensions span three signal families: \textit{semantic
understanding}, \textit{execution quality}, and \textit{reliability}. Question
interpretation and SQL and data sourcing form the agent's semantic
understanding. Insight reasoning and output quality form its execution
quality. Cross-run repeatability across all four dimensions is its
reliability.

These dimensions imply three evaluation requirements drawn from two
complementary views. The analytics view asks whether the agent selected the
right metric, source, grain, and interpretation. The process view asks whether
it used tools correctly, recovered from errors, avoided loops, and escalated
when evidence was weak. Combining both views keeps the evaluation from hiding
either domain mistakes or trajectory mistakes.

First, evaluation must be trajectory-aware. Tool-using failures are often
invisible in the final answer. The agent may call the right tool with wrong
arguments. It may skip a prerequisite lookup, loop over schema discovery, fail
to recover, or avoid escalation despite weak evidence.

Second, evaluation must be analytics-specific. Generic trajectory checks do not
know whether a table is canonical. They also do not know whether a metric has a
local definition or a required filter. Analytics agent evaluation must map
generic tool failures into domain failures. Examples include wrong source,
wrong grain, wrong time window, missing decomposition, and unsupported guidance.

Third, evaluation must measure repeatability. A single successful run does not
establish reliability. If the same question produces different tables, numbers,
or conclusions across repetitions, the agent may create a false sense of
stability. Repeatability is especially important for high-stakes analytics
questions such as forecasting, anomaly explanation, experiment readouts, and
executive reporting.

\section{End-to-End Evaluation Methodology}
\label{sec:methodology}

\begin{figure*}[t]
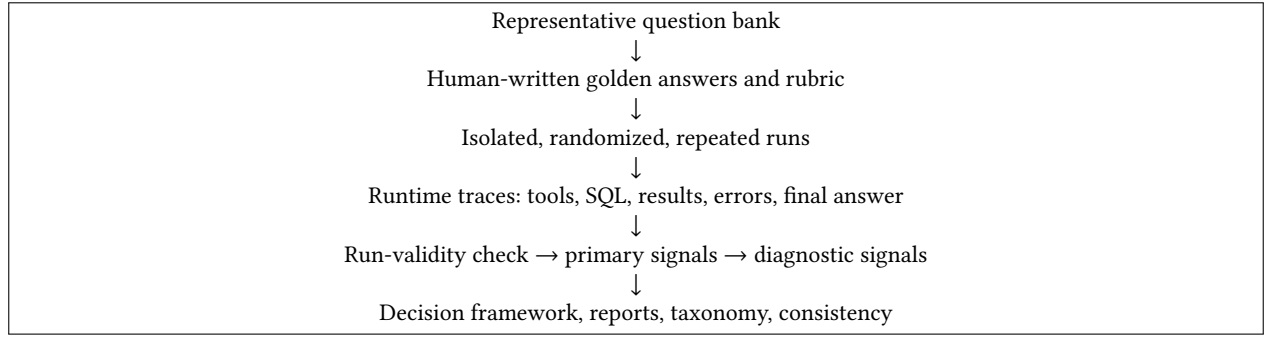

\centering
\fbox{\begin{minipage}{0.92\linewidth}
\centering
Representative question bank\\
$\downarrow$\\
Human-written golden answers and rubric\\
$\downarrow$\\
Isolated, randomized, repeated runs\\
$\downarrow$\\
Runtime traces: tools, SQL, results, errors, final answer\\
$\downarrow$\\
Run-validity check $\rightarrow$ primary signals $\rightarrow$ diagnostic signals\\
$\downarrow$\\
Decision framework, reports, taxonomy, consistency
\end{minipage}}
\caption{End-to-end evaluation pipeline for analytics agents.}
\Description{A pipeline from question bank and golden answers to repeated runs, trace capture, validation, scoring, reports, taxonomy, and consistency analysis.}
\label{fig:pipeline}
\end{figure*}

The methodology runs in three stages. First, the evaluation builds a
representative question bank with human-written golden answers and runs the
agent in isolated, randomized, repeated trials, capturing runtime traces.
Second, each run passes a run-validity check, then receives tiered scoring
that separates primary signals from diagnostic signals and treats abstention
as a first-class verdict. Third, the scores feed a decision framework
calibrated against domain-expert review. Figure~\ref{fig:pipeline} summarizes
the pipeline; the subsections that follow describe each stage.

\subsection{Representative Question Bank}

The evaluation begins with a question bank sampled from realistic analytics
workflows. Questions should be tagged by business area, question type, source
tier, ambiguity, and decision risk. This prevents one aggregate score from
mixing exploratory trend questions with high-precision finance or forecast
questions that require stricter reliability.

\subsection{Human-Written Golden Answers}

Golden answers specify the expected analytical contract. They may include
required sources, prohibited sources, date windows, and filters. They may also
include decomposition steps, numeric bands, expected terms, and claims the
answer must not make. A golden answer is not merely a final answer. It is a
compact description of the workflow that a domain expert expects. In the case study, each golden answer was drafted by the domain expert who
owns that business area and independently reviewed and approved by a second analyst, tracked by an explicit draft-versus-approved status flag; we did not compute a formal inter-rater reliability statistic (Section~\ref{sec:limitations}).

Golden answers also bridge semantic knowledge and evaluation. A future semantic
context layer may move the same knowledge into runtime guidance. That guidance
can include metric definitions, source hierarchy, canonical queries, business
rules, and decomposition templates. These fields define what runtime semantic
guidance would need to cover.

\subsection{Isolated Repeated Runs}

Each question is run in an isolated session. The runner resets state between
questions and randomizes question order with fixed seeds. It also prompts the
agent to treat each question independently. Repetition makes stochastic behavior
observable. The evaluation does not ask whether one sampled answer was good. It
asks whether the agent repeatedly chooses the same interpretation.

\subsection{Runtime Trace Capture}

For every run, the evaluation records the raw question and configuration
metadata. It also records tool calls, query text, query outputs, errors,
retries, latency, tool-round counts, and final answer. These traces are the
central evidence object. If a result is wrong, the trace can localize the
failure. It can point to interpretation, source selection, SQL construction,
tool failure, result interpretation, or final answer synthesis.

\subsection{Run Validity Before Scoring}

Before scoring answer quality, the evaluation checks whether a run is valid
enough to interpret. The case study used validation signals for hard errors and
early refusal patterns. It also checked user-visible authentication mentions,
real-data presence, and infrastructure authentication failures in tool outputs.
This check prevents a broken execution environment from being mistaken for agent
behavior.

\subsection{Primary, Diagnostic, and Abstention-Aware Scoring}
\label{sec:scoring}

The methodology separates primary signals from diagnostic signals. Primary
signals assess the core analytical contract for review, comparison, or
intervention measurement. Diagnostic signals explain why a primary result
changed. This distinction matters because analytics evals can easily accumulate
many scorers; reviewers should not have to interpret every diagnostic as a
release gate.

Every scorer returns one of four verdicts: \textit{yes}, \textit{partial},
\textit{no}, or \textit{unknown}. The \textit{unknown} verdict is a
first-class abstention, not a failure. Aggregates should report coverage
separately from pass rate so that missing golden answers, questions outside
scope, and unavailable trace fields are visible. Concretely, unknown verdicts
are excluded from the denominator and reported separately as a coverage rate;
among the decided runs, each verdict is scored yes $=$ 1.0, partial $=$ 0.5,
no $=$ 0.0, and the reported pass rate is the mean of these scores. Binary
scorers that never emit \textit{partial} reduce to yes / (yes $+$ no).

\subsection{From Scores to Decisions}

Primary and diagnostic scores support engineering and review decisions. They
do not define a universal release policy. Table~\ref{tab:decision-framework}
shows how this paper interprets the signals. A run-validity failure means that
quality scores should not be trusted. Primary signals compare versions or
interventions. Diagnostic signals identify repair work. Cross-run checks test
interpretation stability. Human calibration is required before automated judges
or thresholds become deployment gates.

\begin{table*}[t]
\caption{Decision framework for interpreting evaluation signals.}
\label{tab:decision-framework}
\small
\begin{tabular}{@{}p{0.18\textwidth}p{0.25\textwidth}p{0.27\textwidth}p{0.23\textwidth}@{}}
\toprule
Eval layer & Question answered & Example signals & Decision use \\
\midrule
Run-validity check & Was the run environment valid enough to score? & Hard errors, real-data rate, early-refusal rate, infrastructure-auth errors & Discard or rerun invalid batches before interpreting quality \\
Primary correctness & Did the agent satisfy the core analytical contract? & Required source, answer attempt, escalation behavior, key numeric bands & Compare model, prompt, tool, or intervention variants \\
Diagnostic signals & Why did the result fail or improve? & Schema overrun, execution efficiency, self-consistency, retry recovery & Choose the next repair target \\
Cross-run reliability & Does the same question produce the same interpretation? & Stable, value-unstable, interpretation-unstable, no-attempt labels & Route unstable patterns to human review or stricter workflow control \\
Human calibration & Do automated judgments match expert review? & Expert rubric grades, judge agreement, threshold review & Calibrate thresholds before they become deployment gates \\
\bottomrule
\end{tabular}
\end{table*}

We therefore call this a decision framework rather than a validated release
policy. The method makes failures visible and actionable. Exact thresholds
should be calibrated with domain-expert review before they become blocking
deployment criteria.

\subsection{Human Calibration}

Automated scoring should be calibrated against human review before it drives
blocking or approval decisions. Domain experts can grade a sample of traced runs
using the four-dimension rubric, then compare automated verdicts to expert
judgments. In the case study, deterministic and trace-backed signals form the
main evidence; calibrated LLM judges remain a follow-up direction.

\section{Analytics-Specific Failure Taxonomy}

Table~\ref{tab:taxonomy} summarizes the taxonomy used to interpret results and
prioritize interventions. The taxonomy is deliberately workflow-oriented:
each failure mode points to a potential repair in context, tools, workflow
enforcement, or review policy.

The taxonomy operationalizes both views from Section~\ref{sec:why}. Question interpretation,
semantic/source selection, analytical method, and insight reasoning capture
analytics-specific correctness. SQL and tool execution, confidence and
grounding, and repeatability capture process failures in tool-using agents.

\begin{table*}[t]
\caption{Analytics-agent failure taxonomy.}
\label{tab:taxonomy}
\begin{tabular}{@{}p{0.19\textwidth}p{0.38\textwidth}p{0.36\textwidth}@{}}
\toprule
Layer & Failure mode & Typical intervention \\
\midrule
Question interpretation & Wrong metric, scope, segment, or ambiguity handling & Clarification policy, intent parser, examples \\
Semantic/source selection & Wrong source of truth or metric definition & Source hierarchy, metric glossary, canonical routing \\
SQL and tool execution & Wrong time window, grain, join, filter, or query recovery & Schema lookup, validation checks, row-count checks \\
Analytical method & Missing decomposition or diagnostic workflow & Domain workflow templates, required intermediate steps \\
Insight reasoning & Unsupported causal or decision claim & Alternative-hypothesis checks, escalation gates \\
Confidence and grounding & Plausible but wrong numeric answer or contradiction & Numeric calibration, SQL-answer grounding checks \\
Repeatability & Cross-run table, value, or narrative instability & Repeated-run checks, deterministic planning, canonical routes \\
\bottomrule
\end{tabular}
\end{table*}

The taxonomy also clarifies why semantic context alone is insufficient.
Guidance tells the model what good behavior looks like. Workflow enforcement
prevents the model from skipping required steps. Trajectory evaluation verifies
whether those steps happened. Analytics agents need all three.
% [AUDIT 2026-06-10, Shishir review — S-2: Contribution 2 claims the taxonomy
% "connects generic trajectory failures to concrete data and decision risks,"
% but the Results section never tags an observed failure to its taxonomy layer.
% Suggest adding one clause per Results subsection (or a taxonomy-layer column)
% mapping each finding to a layer, e.g. the YoY contract miss (Sec 6.3) -> the
% Analytical-method layer. Closes the contribution-support gap. Teja's call.]

\section{Case Study}
\label{sec:case}

We evaluated a controlled internal analytics agent at a large online
marketplace. The agent answers natural-language business questions using schema
lookup and query execution tools over an enterprise warehouse, then synthesizes
a natural-language answer. For anonymity, we refer to the two model
configurations as \textbf{Fast} and \textbf{Reasoning}.

The case study used a 50-question bank balanced across five internal
business areas, with 10 questions per area. Each question ran three times
against each of two model configurations, yielding 300 total traces. Both
configurations used the same toolset, prompt scaffold, and runner. The
intended difference was the model configuration.

Finance was the one area with complete structured golden answers. The case
study uses those 10 finance questions for the structured-golden scorecard.
We score them across the same three randomized repetitions per configuration,
producing 30 finance traces per configuration. The other four areas lack
equivalent golden coverage. We therefore treat the finance scorecard as
diagnostic evidence, not as a complete benchmark.

\section{Results}

The results are not presented as a model benchmark. They are evidence that
end-to-end evaluation exposes failures hidden by grading final answers. The
five subsections give five views of the same 300 traces: run validity, the
finance scorecard, contract violations, trace-only risks, and cross-run
consistency.

\subsection{Run Validity and Answering Behavior}

\begin{table}[ht]
\caption{Top-level run metrics across 150 traces per configuration.}
\label{tab:validity}
\begin{tabular}{lrr}
\toprule
Metric & Fast & Reasoning \\
\midrule
Traces per configuration & 150 & 150 \\
Hard errors & 1 / 150 & 0 / 150 \\
Hit tool-round limit & 0 / 150 & 24 / 150 \\
Early refusal rate & 73\% & 0\% \\
Real-data rate & 21\% & 73\% \\
Average latency & 41 s & 119 s \\
p90 latency & 86 s & 234 s \\
Run-validity check & Fail & Pass \\
\bottomrule
\end{tabular}
\end{table}

The Reasoning configuration answered far more often, as
Table~\ref{tab:validity} shows. It reduced early refusal from 73\% to 0\% and
increased real-data answers from 21\% to 73\%. But it was also much slower.
It hit the maximum tool-round budget on 16\% of runs. The dominant failure
mode changed from early abandonment to over-exploration and unstable
interpretation.

\subsection{Finance Golden Answers}

\begin{table}[ht]
\caption{Selected finance scorer deltas. Pass rate is the mean verdict credit over decided runs, with unknowns excluded and reported as coverage (Section~\ref{sec:scoring}); row populations vary by scorer applicability. ``Should not decline'' captures cases where the agent declined despite sufficient evidence; ``Decomposition method'' captures whether the answer applied the expected analytical decomposition.}
\label{tab:scorecard}
\begin{tabular}{lrrr}
\toprule
Scorer & Fast & Reasoning & Delta \\
\midrule
Escalation discipline & 29\% & 93\% & +64 pp \\
Required table used & 0\% & 50\% & +50 pp \\
Should not decline & 65\% & 100\% & +35 pp \\
Revenue YoY band & 62\% & 83\% & +21 pp \\
Decomposition method & 3\% & 10\% & +7 pp \\
Execution efficiency & 74.2\% & 6.7\% & -67.5 pp \\
Prohibited table avoided & 100\% & 100\% & 0 pp \\
Retry recovery & 100\% & 100\% & 0 pp \\
\bottomrule
\end{tabular}
\end{table}

Table~\ref{tab:scorecard} reports selected finance scorer deltas. The
Reasoning configuration improved several primary signals: it attempted
questions more often, escalated more appropriately, and found required source
tables more often. Yet required-table use reached only 50\%, and the
decomposition method remained weak for both configurations. Better model
capability improved willingness and exploration, but did not reliably produce
the expected analytical method.

\subsection{Correct Number, Wrong Analytical Contract}

One finance question asked for a year-over-year Q1 revenue comparison. In a
representative Reasoning run, the agent selected a plausible revenue table.
Headline revenue numbers fell within about 0.3\% of the golden answer. Monthly trends were also close.

However, the answer failed important parts of the analytical contract. It did
not use the prescribed source of truth table. It did not compute the required
projects and revenue per project decomposition. It also omitted expected driver
checks. A final answer evaluator might reward the answer because the headline
number was close. An end-to-end evaluator flags it as incomplete because it
skipped the method required to explain why revenue moved. This divergence is not rare. Across the 30 Reasoning finance traces, a
final-answer grader sees a completed, plausible answer on every trace, yet
trace-backed scoring shows a non-canonical source-of-truth table on 15 of 30
(50\%) and a skipped decomposition on 27 of 30 (90\%). Final-answer-only
grading would therefore still rank Reasoning above Fast, but it would report
the upgrade as an unqualified win and hide these contract failures.

\subsection{Trace-Only Risks}

Trace-only diagnostics expose risks that do not require a complete golden
answer. Table~\ref{tab:traceonly} reports five such signals across 150 traces
per configuration. The Reasoning configuration produced more implausible
numeric claims and more hard contradictions than the Fast configuration. Fast
looks safer at first glance, but its lower numeric-claim rate is partly an
artifact of answering fewer questions with numerics. Silence is not reliability.

\begin{table}[ht]
\caption{Trace-only diagnostic signals across 150 traces per configuration. Counts are over 150 traces; percentages are out of 150.}
\label{tab:traceonly}
\begin{tabular}{lrr}
\toprule
Diagnostic & Fast & Reasoning \\
\midrule
Schema lookup calls (total)            & 115        & 961         \\
Schema lookups per trace (avg)         & 0.77       & 6.4         \\
Schema overrun rate                    & 1\%        & 77\%        \\
Implausible numeric claims             & 3 (2.0\%)  & 16 (10.7\%) \\
Hard SQL-answer contradictions         & 8 (5.3\%)  & 11 (7.3\%)  \\
\bottomrule
\end{tabular}
\end{table}

\subsection{Cross-Run Consistency}

\begin{table}[ht]
\caption{Cross-run determinism across 50 questions.}
\label{tab:determinism}
\begin{tabular}{lrrrrr}
\toprule
Model & Stable & Value & Interpretation & Partial & No attempt \\
\midrule
Fast & 0 & 19 & 21 & 4 & 6 \\
Reasoning & 1 & 8 & 41 & 0 & 0 \\
\bottomrule
\end{tabular}
\end{table}

Table~\ref{tab:determinism} shows the cross-run determinism breakdown. The
strongest reliability signal was interpretation instability. The Reasoning
configuration attempted every question. But on 41 of 50 questions, it changed
the table set or query interpretation across repetitions. This is a
decision-grade reliability risk. A user may receive a coherent answer today
and a different coherent answer tomorrow. The answer may not show that the
underlying source of truth changed.

\section{From Evaluation to Intervention}

The case study supports three lessons that connect the methodology back to
intervention strategy. Each lesson points to a specific class of intervention.
Section~\ref{sec:governance} names the governance primitives those
interventions require before any answer can be trusted for a business
decision.

\textbf{Model upgrades are not enough.} The Reasoning configuration answered
more questions and produced more grounded outputs. It still had source of truth
errors, weak decomposition, schema overrun, contradictions, and interpretation
drift. Model capability changed the failure surface; it did not remove the need
for workflow control.

\textbf{Analytics agents need semantic context plus enforcement.} The observed
failures point to interventions such as metric glossaries, source hierarchy,
canonical table routing, domain decomposition templates, row-count checks, and
post-query validation. But guidance must be paired with enforcement. The agent
must do more than access definitions or canonical patterns. The workflow must
make lookup, validation, and escalation observable and testable.

\textbf{Evaluation should guide intervention measurement.} Broad interventions
such as a semantic context layer should be decomposed where possible. Smaller
tests can target table metadata, metric definitions, source hierarchy, query
patterns, and workflow scaffolds. When decomposition is not possible, per-layer
scoring can still show what changed. A semantic intervention might improve
source selection without improving answer completeness. That is still
actionable.

\subsection{Governance for Business Decisions}
\label{sec:governance}

In this paper, governance means the evaluation primitives needed before an
analytics-agent answer can be trusted for business decisions. The methodology
provides five such primitives, each grounded in a Section~\ref{sec:methodology} mechanism:

\begin{enumerate}
  \item \textbf{Traceability} via runtime trace capture.
  \item \textbf{Abstention and escalation} via abstention-aware scoring.
  \item \textbf{Primary vs diagnostic roles} via tiered scoring.
  \item \textbf{Run stability} via isolated repeated runs and cross-run consistency analysis.
  \item \textbf{Version comparison} via the trace-backed contract.
\end{enumerate}

These primitives make governance possible; they do not constitute a deployed
governance program.

\section{Related Work}

Agent evaluation increasingly recognizes that final answers are insufficient.
ReAct shows the value of interleaving reasoning and acting, making trajectories
central to agent behavior~\cite{yao2023react}. GAIA and tau-bench evaluate
tool use and multi-step agent reliability beyond simple question
answering~\cite{mialon2023gaia,yao2024taubench}. Tau-bench is especially
relevant because it evaluates repeated task success and consistency across
trials in tool-using domains.

Text-to-SQL work such as Spider, BIRD, and Spider 2.0 evaluates natural
language to database-query capability~\cite{yu2018spider,li2023bird,spider2024}.
These benchmarks are essential, but enterprise analytics agents add
requirements beyond SQL execution. The agent must choose among competing
source of truth tables, respect local business definitions, and explain results
according to domain practice.

LLM-as-a-judge work, including MT-Bench and pairwise arena benchmarks, studies
when model judges agree with humans~\cite{zheng2023judge}. Self-refinement
methods show how LLMs can critique and improve their own
outputs~\cite{madaan2023selfrefine}.
Our approach is compatible with judge-based scoring, but the reported evidence
relies primarily on deterministic and trace-backed signals. LLM judges are most
useful for nuanced interpretation and insight quality after calibration against
domain-expert grades.

Enterprise analytics evaluation differs from public benchmarks. Governed
warehouses, local metric definitions, and source-of-truth hierarchies make
analytics-agent correctness domain-specific in ways that public text-to-SQL
or trajectory benchmarks cannot capture. The methodology in this paper adds
three contributions not covered by the prior work above. First, a
workflow-level evaluation ties trajectory checks to analytical contracts.
Second, an analytics-specific failure taxonomy maps generic tool failures to
domain repair targets. Third, a 300-trace case study demonstrates the
methodology and exposes failure modes that final-answer evaluation hides.

\section{Limitations and Future Work}
\label{sec:limitations}

This study has several limitations. First, the data and agent are internal to
one company. Second, the case study compares two model configurations, but it
is not a full model benchmark. Third, the case study did not evaluate a
deployed semantic layer. The evidence shows where semantic context and
canonical routing are needed but does not show that they already improved the
agent.
% [AUDIT 2026-06-10, Shishir review — added caveats S-1/S-3/N1; Teja please accept/reject]
Fourth, cross-run consistency is estimated from three repetitions per question;
three samples bound but do not precisely estimate run-to-run stability, so the
interpretation-instability counts should be read as lower bounds rather than
exact rates. Fifth, latency and execution-efficiency figures are specific to
this data warehouse and execution environment; they reflect the deployment, not
intrinsic model properties, and should not be read as portable model
benchmarks. Sixth, the finance golden answers were authored internally without a
reported inter-rater reliability check, and the two model configurations are
anonymized as ``Fast'' and ``Reasoning,'' which limits how far a reader can
attribute the observed deltas to specific model families. Authoring and reviewing a structured golden answer took on the order of a few
hours of a domain-familiar data scientist per question, dominated by encoding
the analytical contract and the second-analyst review; reducing this cost
through calibrated automation of the more mechanical checks is future work.

Future work has three priorities. First, calibrate the decision framework
against domain-expert review so that primary-signal thresholds become
trustworthy gates. Second, instrument a semantic context layer (metric
glossary, source hierarchy, canonical routing) and measure each component
against the same trace-backed contract. Third, apply the methodology to a
second analytics agent on a different domain to test how the failure taxonomy
generalizes.

\textbf{Reproducibility.} The methodology, scorers, and decision framework
are independently reproducible. Practitioners can apply them to their own
analytics agents, question banks, and golden answers. The specific case
study is not directly reproducible. The question bank, the agent system,
and the data warehouse are internal to one organization and cannot be
shared. The portable contribution is the methodology, not the specific
rates.

\section{Conclusion}

Analytics agents need evaluation at the level of the analytical workflow, not
only the final answer. The methodology combines question banks, human golden
answers, repeated runs, and runtime traces. Each run passes a run-validity
check before tiered, primary and diagnostic scoring. It also adds
abstention-aware scoring, a decision framework, and cross-run consistency
analysis.

The case study shows why that end-to-end view matters. A higher-capability
configuration answered more questions and used real data more often. It still
produced source of truth errors, weak decomposition, schema overrun,
contradictions, and interpretation drift. These are exactly the failures that
final answer evaluation hides. Trace-backed, end-to-end evaluation gives teams
the measurement surface needed to improve analytics agents without mistaking
plausibility for reliability.

\bibliographystyle{ACM-Reference-Format}
\bibliography{analytics_agent_eval_kdd2026_refs}

\end{document}